\documentclass{article}

\usepackage{PRIMEarxiv}

\usepackage[utf8]{inputenc} 
\usepackage[T1]{fontenc}    
\usepackage{hyperref}       
\usepackage{url}            
\usepackage{multirow, multicol, makecell, booktabs}       
\usepackage{amsfonts}       
\usepackage{nicefrac}       
\usepackage{microtype}      
\usepackage{lipsum}
\usepackage{fancyhdr}       
\usepackage[sort]{cite}
\usepackage{graphicx}       
\graphicspath{{media/}}     
\usepackage{float}

\usepackage{amsfonts,amsmath,amssymb,bm}
\let\vec\bm
\usepackage[per-mode=fraction,exponent-product=\cdot]{siunitx}
\usepackage{booktabs, xcolor}
\usepackage{tabularx}
\usepackage{hyperref}
\hypersetup{%
  colorlinks=false,
  linkbordercolor=black,
  pdfborderstyle={/S/U/W 1}
}

\usepackage{subfigure}
\usepackage{overpic}
\newcolumntype{C}{>{\centering\arraybackslash}X}

\usepackage{tikz}
\usetikzlibrary{positioning, arrows.meta, shapes.geometric}

\title{Feasibility study on solving the\\Helmholtz equation in 3D with PINNs} 

\author{ \href{https://orcid.org/0000-0002-2148-6703}{\includegraphics[scale=0.06]{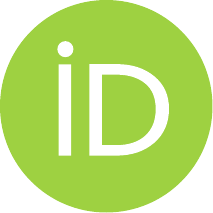}\hspace{1mm}Stefan~Schoder}\\
	Institute of Fundamentals and Theory in Electical Engineering (IGTE)\\
    Graz University of Technology\\
	8010 Graz, Austria \\
	\texttt{stefan.schoder@tugraz.at} \\
	 \And
	 \href{https://orcid.org/0000-0002-8652-5048}{\includegraphics[scale=0.06]{orcid.pdf}\hspace{1mm}Florian  Kraxberger} \\
    Institute of Fundamentals and Theory in Electical Engineering (IGTE)\\
    Graz University of Technology\\
	8010 Graz, Austria \\
 \texttt{kraxberger@tugraz.at} \\
}

\begin{document}
\maketitle

\begin{abstract}
Room acoustic simulations at low frequencies often face significant uncertainties of material parameters and boundary conditions due to absorbing material. We discuss the application of Physics-Informed Neural Networks (PINNs) to solve the (forward) Helmholtz equation in three dimensions (3D), employing mini-batch stochastic gradient descent with periodic resampling every 100 iterations for memory-efficient training. Addressing the computational challenges posed by the extension of PINNs from 2D to 3D for acoustics, DeepXDE is used for implementing the forward PINN. The proposed numerical method is benchmarked against an analytical solution of a standing wave field in 3D. The PINN results are also compared to the Finite Element Method (FEM) solutions for a 3D wave field computed with openCFS. The alignment between PINN-generated solutions and analytical/FEM solutions shows the feasibility of PINNs modeling 3D acoustic applications for future inverse problems, and validating the accuracy and reliability of the proposed approach. 
Compared to FEM, establishing the PINN model took few hours (similar to the setup of a FEM simulation), the training took 38h to 42.8h (which is longer than the solution of the FEM simulation, which took 17min-19min), and the inference took 0.05 seconds being more than 20,000 times faster than the FEM benchmark openCFS using the same number of degrees of freedomwhen producing the results. Thereby, the insight is gained that 3D acoustic wave simulations in the frequency domain are feasible for forward PINNs and can predict complex wave behaviors in real-world applications.
\end{abstract}

\keywords{PINNs \and FEM \and Helmholtz equation \and DeepXDE \and openCFS \and Acoustics \and Waves \and 3D \and PDE \and Room Acoustics \and Time-Harmonic Wave Field \and Absorber}

\section{Introduction}

Room acoustics simulations aim towards improving the acoustic performance of interior spaces by meeting specific criteria and predicting acoustic quality in advance of the construction phase. However, numerous challenges have to be overcome, including uncertain boundary conditions and the simulation of low-frequent wave phenomena \cite{schmid4545338physics,Kraxberger2023Validated}. A comprehensive examination underscores the current obstacles, with a particular emphasis on the persisting challenge of errors arising from uncertain input parameters. A survey identified that about two-thirds of the acousticians judge uncertainties in input parameters as the foremost obstacle hindering more accurate simulation results \cite{jeong2022room}. In \cite{vorlander2013computer}, simulations with the use of absorption coefficients from textbook tables showed inadequate prediction capabilities of reverberation time. These insights underscore the critical importance of using correct material data for the accurate modeling of room acoustics simulations, manifesting the interest in using forward physics-informed neural networks (PINNs) and inverse PINNs in three-dimensional (3D) space \cite{schmid4545338physics}. The computation of the time-harmonic acoustic wave field in large 3D geometries represents a challenging frontier for numerical methods. Especially, the solutions to the Helmholtz equation in 3D are essential for understanding low-frequent room acoustic scenarios. Traditional numerical methods, such as Finite Element Methods (FEM), have provided valuable insights \cite{Kraxberger2023Finite,Kraxberger2023Nonlinear}, but their computational approach is limited, and the inverse estimation of parameters is challenging. In this context, PINNs show a promising supplement, leveraging the power of optimization and machine learning to tackle both the forward propagation of acoustics and the inverse estimation of material parameters of 3D acoustic fields.

This paper addresses the challenge associated by extending PINNs to solve the Helmholtz equation in 3D, presenting a methodological breakthrough facilitated by the integration of state-of-the-art mini-batch stochastic gradient descent optimization with periodic shuffling of the training data points. 
The PINN investigation is conducted within the DeepXDE framework \cite{lu2021deepxde} and compared to a continuously tested and validated \cite{schoder2020hybrid,schoder2020computational,schoder2020aeroacoustic,schoder2022aeroacoustic,tieghi2023machine,schoder2022error,maurerlehner2022aeroacoustic,schoder2023acoustic,schoder2023dataset,wurzinger2024experimental,Kraxberger2023Validated} FEM solver openCFS \cite{Schoder2022openCFS}. 
Physics-Informed Neural Networks offer a unique advantage, using the same optimization routines when establishing the forward PINN and the inverse PINN. By incorporating domain-specific knowledge and physical principles into the neural network architecture, PINNs have demonstrated a capacity to learn and predict complex solutions to partial differential equations \cite{raissi2019physics, Schoder2023Posta}. This study builds upon the success of PINNs in 1D and 2D acoustic simulations \cite{schmid4545338physics,lu2021deepxde,Song2021Solving,EscapilInchauspe2023Hyper,Gladstone2022FO,Wu2023Helmholtz} (see Tab.~\ref{tab:summary}, extending their application to the more challenging 3D acoustics problems. The goal is to model and predict wave phenomena in 3D space accurately.

\begin{table}[h]
    \centering
    \caption{Literature overview on noteworthy public development projects solving acoustic fields.}
    \begin{tabularx}{\textwidth}{@{}p{2cm}Cp{1cm}p{1cm}@{}}
    \toprule
        \textbf{Type} & \textbf{Name} & \textbf{URL} & \textbf{Ref.} \\
\midrule        
Article, GIT & DeepXDE: A Deep Learning Library for Solving Differential Equations & \href{https://github.com/lululxvi/deepxde}{GIT}  &   \cite{lu2021deepxde} \\
        Article, GIT & Solving the frequency-domain acoustic VTI wave equation using physics-informed neural networks & \href{https://github.com/swag-kaust/PINN-Helmholtz-solver}{GIT} &  \cite{Song2021Solving} \\
        Article & Hyper-parameter tuning of physics-informed neural networks: Application to Helmholtz problems &  &  \cite{EscapilInchauspe2023Hyper} \\
        Preprint & FO-PINNs: A First-Order formulation for Physics Informed Neural Networks &  &  \cite{Gladstone2022FO} \\
        Article & Helmholtz-equation solution in nonsmooth media by a physics-informed neural network incorporating quadratic terms and a perfectly matching layer condition &  &  \cite{Wu2023Helmholtz} \\
        Preprint & PINNs-TF2: Fast and User-Friendly Physics-Informed Neural Networks in TensorFlow V2 & \href{https://github.com/rezaakb/pinns-tf2}{GIT} &  \cite{Bafghi2023PINNs} \\
        Preprint & Physics-Informed Neural Networks for Acoustic Boundary Admittance Estimation & \href{https://papers.ssrn.com/sol3/papers.cfm?abstract_id=4545338}{Preprint} &  \cite{Schmid2023Physics} \\
         \bottomrule
    \end{tabularx}
    \label{tab:summary}
\end{table}

The article is structured as follows: Sec. \ref{sec:PINN} introduces the theoretical background of PINNs as self-supervised learning. Section \ref{sec:examples} presents the numerical examples of 3D room-like geometry with different boundary conditions, differential equation forcing, and material parameters. This section also shows the training results and the comparison to the analytic solution and FEM solution by error analysis. Each numerical example is accompanied by a discussion of the results obtained. Finally, Sec. \ref{sec:conclusion} summarizes the key findings and provides a roadmap for future research directions.

\section{Physics-informed Neural Networks for the Helmholtz Equation in Acoustics} \label{sec:PINN}

\begin{figure}
    \centering
    \includegraphics[width=0.8\textwidth]{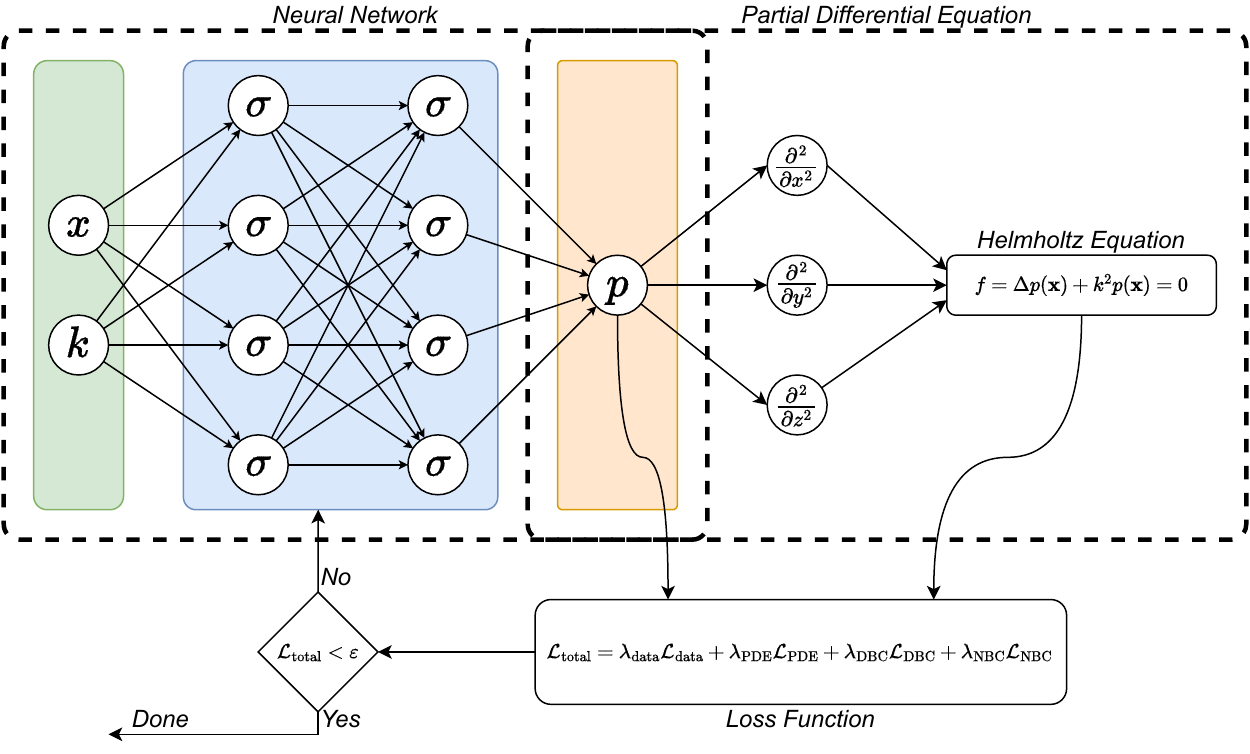}
    \caption{PINNs applied to Helmholtz equation.}
    \label{fig:PINN}
\end{figure}

PINNs represent a cutting-edge fusion of neural networks and equation-based principles of physics, offering a bundle of possibilities for solving complex partial differential equations (PDEs) and simulating physical systems \cite{raissi2019physics}. This innovative approach delegates implementation-associated challenges encountered in traditional computational methods, such as Finite Element Methods (FEM) or Finite Difference Methods, by seamlessly integrating computational domain-specific and mathematical equation-specific knowledge directly into the neural network architecture. Specifically, Figure~\ref{fig:PINN} schematically shows how such a feed-forward (multi-layer perceptron \cite{rosenblatt1958perceptron,haykin1998neural}) fully connected layer network is tailored to model the inhomogeneous Helmholtz equation \begin{equation}
    (\Delta + k^2) p(\omega, \bm x) = f(\omega, \bm x)
    \label{eq:Helmholtz}
\end{equation} 
where \(p \in \mathbb{C}^3\) is the time-harmonic acoustic pressure, \({k} = \omega/c\) is the wavenumber, $\omega$ the angular frequency, $c$ the speed of sound, $f$ the forcing term, and \(\Delta\) is the Laplacian operator.

To predict $\hat{p}(\omega, \bm x)$ with a PINN, let $\mathcal{N}^L(\mathbf{x}): \mathbb{R}^{d_{\text {in }}} \rightarrow \mathbb{R}^{d_{\text {out }}}$ be a $L$-feed-forward neural network (FNN) (with $(L-1)$-hidden neural network layers), with $N_{\ell}$ neurons in the $\ell$-th layer $\left(N_0=d_{\text {in }}, N_L=d_{\text {out }}\right)$. The output at layer of a feed-forward neural network is calculated by repeated nonlinear activation-function $\sigma$ weighted tensor products of (i) inputs and (ii) bias vector and weight matrix (e.g. in the $\ell$-th layer by $\boldsymbol{W}^{\ell} \in \mathbb{R}^{N_{\ell} \times N_{\ell-1}}$ and $\mathbf{b}^{\ell} \in \mathbb{R}^{N_{\ell}}$, respectively), such that the FNN \footnote{Note that all the network parameters depend on the solution frequency $\omega$. When the frequency of the problem is changing, the FNN must be retrained. This is similar to the FEM, where the solver must construct and invert the system matrices for different frequencies separately.} is recursively defined by
\begin{equation}
    \begin{aligned}
\text { input layer: } & \mathcal{N}^0(\mathbf{x})=\mathbf{x} \in \mathbb{R}^{d_{\mathrm{in}}} \\
\text { hidden layers: } & \mathcal{N}^{\ell}(\mathbf{x})=\sigma\left(\boldsymbol{W}^{\ell} \mathcal{N}^{\ell-1}(\mathbf{x})+\boldsymbol{b}^{\ell}\right) \in \mathbb{R}^{N_{\ell}}, \quad \text { for } \quad 1 \leq \ell \leq L-1, \\
\text { output layer: } & \mathcal{N}^L(\mathbf{x})=\boldsymbol{W}^L \mathcal{N}^{L-1}(\mathbf{x})+\boldsymbol{b}^L \in \mathbb{R}^{d_{\text {out }}} ;
\end{aligned}
\end{equation}

In the case of PINNs in 3D, $\bm x = (x,y,z)$ at the input layer, where $(x,y,z)$ denotes the space directions of one point.
The output prediction at the output layer $L$ is calculated by
\begin{align}
    \hat{p} =& \boldsymbol{W}^L \mathcal{N}^{L-1}(\mathbf{x})+\boldsymbol{b}^L = \boldsymbol{W}^L \sigma\left( \bm{W}^{L-1} \mathcal{N}^{L-2}(\mathbf{x})+\boldsymbol{b}^{L-1}\right) +\boldsymbol{b}^L = \dots\\ =& \mathcal{N}^{L} \circ \mathcal{N}^{L-1} \circ \dots \circ \mathcal{N}^{0}\circ \bm x = \mathcal{N}(\bm \theta, \bm x)\, ,
\end{align}
with $\bm{g}$ being a series of mapping functions. The network transformative operations can be summarized by a nonlinear operation $\mathcal{N}(\bm \theta, \bm x)$ and a structure $\bm \theta$ collecting all weights and biases $\bm \theta = \{\boldsymbol{W}^l,\boldsymbol{b}^l \}_{1\leq l \leq L}$. 
During supervised learning, the neural network weights will be learned by labeled training data $\mathcal{D} = \{\bm x_i, p_i\}_{i=1}^{N_L}$ of $N_L$ samples, and unlabeled training data $\mathcal{T} = \{\bm x_i\}_{i=1}^{N_U}$ of $N_U$ samples, to approximate a continuous function mapping $g(\bm x) = p$. $\mathcal{T}$ comprises two sets, the points in the domain $\mathcal{T}_f$ of $N_{PDE}$ samples and the points on the boundary $\mathcal{T}_b$ of $N_{BC}$ samples. The neural network operator $\hat{p} = \mathcal{N}(\bm \theta, \bm x)$ can be interpreted as an approximation of $g(\bm x) = p$, regarding optimality of the discrepancy (error) described by the cost functional (loss function), e.g. based on mean squared error (MSE) for the data loss
\begin{equation}
    \mathcal{L}_\mathrm{data}(\bm \theta,\mathcal{D})  = \frac{1}{N_L} \sum_{i=1}^N \parallel p_i - \hat{p}_i \parallel^2
\end{equation}
with $\|\cdot\|$ being the $L_2$-norm. In general, the data loss can be combined with other types of loss (e.g., boundary loss or PDE loss) as described in sec.~\ref{sec:examples}, resulting in a total loss $\mathcal{L}_\mathrm{total}$. According to the minimization problem
\begin{equation}
\hat{\boldsymbol{\theta}}=\underset{\boldsymbol{\theta}}{\arg \min } \, (\lambda_\mathrm{data}\mathcal{L}_\mathrm{data}(\boldsymbol{\theta},\mathcal{D}) + \lambda_\mathrm{pde}\mathcal{L}_\mathrm{pde}(\boldsymbol{\theta},\mathcal{T}),
\end{equation}
the weights and biases $\boldsymbol{\theta}$ are adjusted to minimize the loss function $\mathcal{L}$. The converged set of weights and biases $\boldsymbol{\theta}$ is the optimal prediction of these parameters, denoted by $\hat{\boldsymbol{\theta}}$. Minimization can be achieved by gradient-based optimization algorithms, like Adam \cite{kingma2014adam} or LBFGS algorithm \cite{liu1989limited}. Using the gradient of the loss function $\nabla_{\boldsymbol{\theta}} \mathcal{L}$, the weights and biases are adapted iteratively to decrease $\mathcal{L}$, typically using backpropagation \cite{rumelhart1986learning} and automatic differentiation \cite{baydin2018automatic}. A standard gradient descent algorithm updates the parameters such that
\begin{equation}
\boldsymbol{\theta}_{i+1}=\boldsymbol{\theta}_i-\alpha_i \nabla_{\boldsymbol{\theta}_i} \mathcal{L} \, ,
\end{equation}
with $\alpha$ representing the learning rate and $i$ the epoch number. 

The Helmholtz equation \eqref{eq:Helmholtz} is a second-order linear partial differential equation of hyperbolic type governing the behavior of acoustic waves in the frequency domain. It is derived from the linear scalar wave equation in the time domain through separation of variables and an assumption of time-harmonic dependence $\tilde{p}(\bm{x},t) = p(\bm{x})e^{-\mathrm{i}\omega t}$. In addition to \eqref{eq:Helmholtz}, for the inhomogeneous Helmholtz equation on a restricted domain $\Omega$, forcing and boundary conditions (e.g., Dirichlet boundary condition at $\Gamma_D$ and Neumann boundary $\Gamma_N$, where the two surfaces form the domain boundary denoted as $\partial \Omega = \Gamma_D \cup \Gamma_N$) have to be considered, forming a well-posed system of equations
\begin{subequations}
\begin{align}
    \Delta p(\bm{x}) + k^2p(\bm{x}) &= f(x), \quad \bm{x} \in \Omega \setminus \partial \Omega \, , \\ 
    p(\bm{x})  &= 0, \quad \bm{x} \in \Gamma_D   \, ,  \\ 
    \nabla p(\bm{x}) \cdot \bm{n} &= 0, \quad \bm{x} \in \Gamma_N   \, . 
\end{align}
\end{subequations}
Here, $f(\bm{x})$ represents the known volume forcing term of the inhomogeneous Helmholtz equation and $\vec{n}$ is the outward pointing normal vector of $\Gamma_N$. The solution to the Helmholtz equation yields the acoustic pressure $p(\bm{x})$, typically expressed as a complex number, at a specific location $\bm{x}$. To incorporate the inherent physical knowledge into the neural network, the residual of the Helmholtz equation $r_\mathrm{PDE}$ is integrated into the loss function.
The residual of the PDE is
\begin{equation}
\begin{split}
    r_\mathrm{PDE}(\bm{x}) &= \Delta \hat{p}(\bm{x}) + k^2\hat{p}(\bm{x}) - f(\bm{x}) \qquad \forall \bm{x} \in D_\mathrm{PDE} \\
    \mathcal{L}_\mathrm{PDE}(\bm \theta,\mathcal{T}_f) &= \frac{1}{N_\mathrm{PDE}} \sum_{i=1}^{N_\mathrm{PDE}} \parallel r_\mathrm{PDE}(\bm{x}) \parallel^2
\end{split}
\label{eq:loss-pde}
\end{equation}
and it is evaluated at randomly sampled collocation points $\bm{x}$ within the computational domain $\Omega$, forming the dataset $\mathcal{T}_{f} = \{ \bm{x}_{i} \}_{i=1}^{N_\mathrm{PDE}}$, where $N_\mathrm{PDE}$ is the number of sample points for the PDE residual. In addition, the residuals of the boundary conditions have to be integrated into the loss function too. The Dirichlet boundary condition residual $r_\mathrm{DBC}$
\begin{equation}
\begin{split}
    r_\mathrm{DBC} &= p(\bm{x}) \qquad \forall \bm{x} \in D_\mathrm{DBC} \, , \\
    \mathcal{L}_\mathrm{DBC}(\bm \theta,\mathcal{T}_{b1}) &= \frac{1}{N_\mathrm{DBC}} \sum_{i=1}^{N_\mathrm{DBC}} \parallel r_\mathrm{DBC}(\bm{x}) \parallel^2
\end{split}
\label{eq:loss-dbc}
\end{equation}
is sampled at the respective sampling points on the Dirichlet boundary $\Gamma_D$ forming the set $\mathcal{T}_{b1} = \{ \bm{x}_{i} \}_{i=1}^{N_\mathrm{DBC}}$, where $N_\mathrm{DBC}$ is the number of sample points for the Dirichlet boundary $\Gamma_D$.
Similarly, the Neumann boundary condition residual $r_\mathrm{NBC}$ is
\begin{equation}
\begin{split}
    r_\mathrm{NBC} &= \nabla p(\bm{x}) \cdot \bm{n}  \qquad \forall \bm{x} \in D_\mathrm{NBC} \, \\
    \mathcal{L}_\mathrm{NBC}(\bm \theta,\mathcal{T}_{b2}) &= \frac{1}{N_\mathrm{NBC}} \sum_{i=1}^{N_\mathrm{NBC}} \parallel r_\mathrm{NBC}(\bm{x}) \parallel^2 \, ,
\end{split}
\label{eq:loss-nbc}
\end{equation}
which is sampled at the respective sampling points on the Neumann boundary $\Gamma_N$ forming the set $\mathcal{T}_{b2} = \{ \bm{x}_{i} \}_{i=1}^{N_\mathrm{NBC}}$, where $N_\mathrm{NBC}$ is the number of sample points for the Neumann boundary $\Gamma_B$.
In the training process, incorporating the BCs involves adding the residual of the BCs directly to the loss function as a mean squared error, corresponding to the forward solution. This results in a total loss
\begin{equation}
    \mathcal{L}_\mathrm{total} =
    \lambda_\mathrm{data}\mathcal{L}_\mathrm{data} + 
    \lambda_\mathrm{PDE} \mathcal{L}_\mathrm{PDE} + 
    \lambda_\mathrm{DBC} \mathcal{L}_\mathrm{DBC} + 
    \lambda_\mathrm{NBC} \mathcal{L}_\mathrm{NBC} \, ,
    \label{eq:loss-total}
\end{equation}
where $\lambda_\mathrm{data}$, $\lambda_\mathrm{PDE}$, $\lambda_\mathrm{DBC}$, and $\lambda_\mathrm{NBC}$ are weighting factors for the individual loss terms. If not defined otherwise $\lambda_\mathrm{PDE}=1$ for all used FNN.

\section{Application} \label{sec:examples}

In this section, basic room application examples in 2D and 3D are presented to highlight the capabilities of PINNs when approximating solutions of the Helmholtz equation.

\subsection{Example 1a - Analytic solution in 2D, Dirichlet} \label{sec:example-1a}

The example demonstrates\footnote{Details of this initial example can be found here \url{https://deepxde.readthedocs.io/en/latest/demos/pinn_forward/helmholtz.2d.dirichlet.html}.} the application of PINNs to solve the Helmholtz equation in a 2D domain with Dirichlet boundary conditions.
The governing equation is eq.~\eqref{eq:Helmholtz} with forcing of 
\begin{equation}
    f(x,y) = k^2 \sin(k x)\sin(k y) \, ,
\end{equation}
with the wave number of $k = 2\pi/\lambda$, wavelength $\lambda=1/2$, and the solution is sought in a 2D spatial domain $\Omega = [0,1]^2$. This leads to the analytic solution 
\begin{equation}
    p(x,y) = \sin(k x)\sin(k y) \, .
\end{equation}

The training process involves formulating a comprehensive loss function that incorporates the Helmholtz equation. For the Helmholtz equation, the loss term is defined in eq.~\eqref{eq:loss-pde}.
A homogeneous Dirichlet boundary condition $p(x,y)=0$ with $(x,y)\in \partial \Omega$ is enforced by a transform of the neural network solution $\mathcal{N}(x,y)$ to the acoustic pressure
\begin{equation}
    \tilde{p}(x,y) = x (x-1) y (y-1) \mathcal{N}(x,y) \, .
\end{equation}
The resulting total loss function for this example is equivalent to eq.~\eqref{eq:loss-total} with $\lambda_\mathrm{data}=\lambda_\mathrm{NBC}=\lambda_\mathrm{DBC}=0$, and therefore reduces to
\begin{equation}
    \mathcal{L}_\mathrm{1a} =
    \mathcal{L}_\mathrm{PDE} \, .
    \label{eq:loss-example1a}
\end{equation}
The PINN is set up using DeepXDE (with the PyTorch backend), and the trained network is capable of predicting the acoustic pressure \(\tilde{p}\) within the specified 2D domain. Ten random collocation points per wavelength along each direction for training and 30 for testing are defined. A fully connected neural network of four layers depth (three hidden layers) and layer width of 150 neurons is used. A sinus activation function is used with Glorot uniform bias and weights initialization. The network is trained over 5000 iterations by the ADAM optimizer with a learning rate of 0.001, resulting in a test loss of 0.00571.\footnote{The defined hyperparameters can be optimized using Ray Tune or scikit-optimize\footnote{\url{https://deepxde.readthedocs.io/en/latest/demos/pinn_forward/helmholtz.2d.dirichlet.hpo.html}}.} 

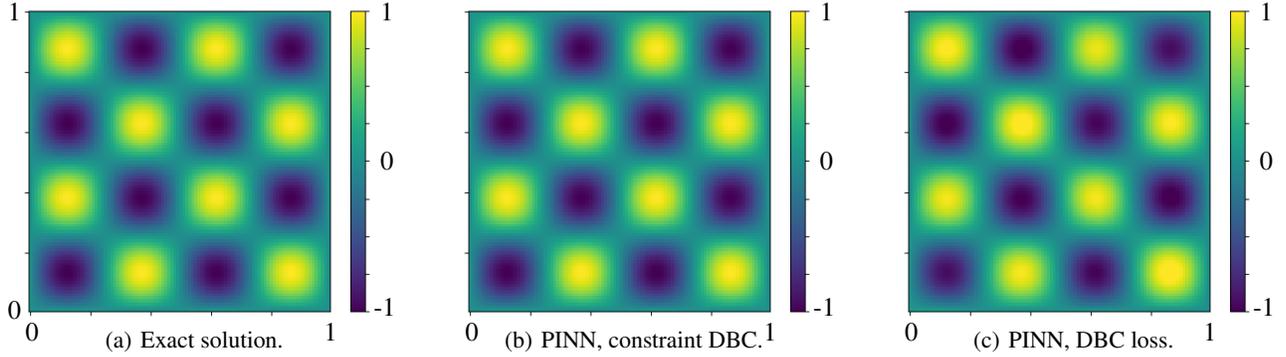
\begin{figure}[ht]
  \centering
  \subfigure[Exact solution.\label{fig:subfig1}]{
  \begin{overpic}[width=0.28\linewidth,trim = 2.5cm 1.1cm 2.9cm 0,clip]
{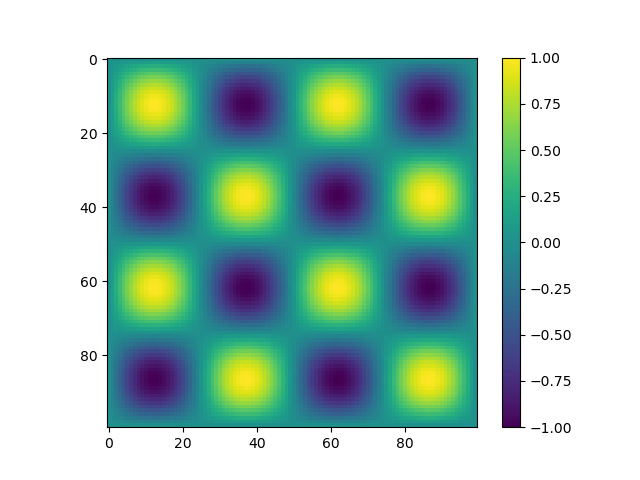}
\put(-4,0){0}
\put(1,-6){0}
\put(-4,84){1}
\put(85,-6){1}
\put(99,1){-1}
\put(99,42){~0}
\put(99,84){~1}
\end{overpic}
  }\hspace{1.0cm}
  \subfigure[PINN, constraint DBC.\label{fig:subfig2}]{
\begin{overpic}[width=0.28\linewidth,trim = 2.5cm 1.1cm 2.9cm 0,clip]
{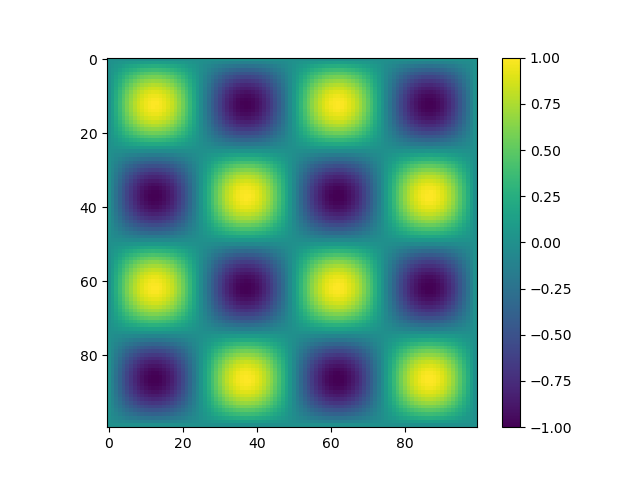}
\put(1,-6){0}
\put(85,-6){1}
\put(99,1){-1}
\put(99,42){~0}
\put(99,84){~1}
\end{overpic}
  }\hspace{1.0cm}
  \subfigure[PINN, DBC loss.\label{fig:subfig3}]{
    \begin{overpic}[width=0.28\linewidth,trim = 2.5cm 1.1cm 2.9cm 0,clip]{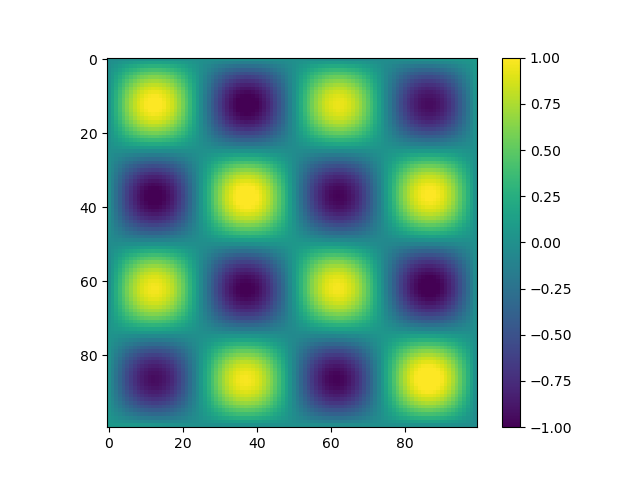}
\put(1,-6){0}
\put(85,-6){1}
\put(99,1){-1}
\put(99,42){~0}
\put(99,84){~1}
\end{overpic}
  }
  \caption{The real part of the acoustic pressure.}
  \label{fig:PINN_2D}
\end{figure}

To enforce the boundary conditions via the loss function using $\lambda_\mathrm{DBC}=100$, 15000 iterations have been made, resulting in a test loss of 0.0761. Figure \ref{fig:PINN_2D} shows the exact (analytical) solution, the PINN solution with constraint DBC, and the PINN solution with DBC modeled as an additional loss term. The figure shows the acoustic pressure of the whole domain.


\subsection{Example 1b - Analytic solution in 3D, Dirichlet} \label{sec:example-1b}
Consistently with the previous example in 3D the computational domain is denoted by $\Omega = [0,1]^3$, the application of PINNs to solve the Helmholtz equation is extended using Neumann boundary conditions. The governing equation is eq.~\eqref{eq:Helmholtz} with forcing 
\begin{equation}
    f(x,y,z) = 2 k^2 \cos(k x)\cos(k y)\cos(k z) \, ,
\end{equation}
with a wave number of $k = 2\pi/\lambda$, wave length $\lambda=1/2$. This leads to an analytic solution of
\begin{equation}
    p(x,y,z) = \cos(k x)\cos(k y)\cos(k z) \, .
\end{equation}
For the Helmholtz equation in 3D, the loss term is defined by eq.~\eqref{eq:loss-pde}.
A homogeneous Dirichlet boundary condition $p(x,y,z)=0$ with $(x,y,z)\in \partial \Omega$ is enforced by a transform of the neural network solution $\mathcal{N}(x,y,z)$ to the acoustic pressure 
\begin{equation}
    \tilde{p}(x,y,z) = x (x-1) y (y-1) z (z-1) \mathcal{N}(x,y,z) \, .
\end{equation}
The resulting total loss function for this example is equivalent to eq.~\eqref{eq:loss-total} with $\lambda_\mathrm{data}=\lambda_\mathrm{NBC}=\lambda_\mathrm{DBC}=0$, and therefore reduces to the one from the previous example \eqref{eq:loss-example1a}. Similarly to the example shown in sec.~\ref{sec:example-1a}, the PINN is set up using DeepXDE (PyTorch backend) with an updated input layer for the additional space domain.

The trained network is capable of predicting the acoustic pressure $\tilde{p}$ within the specified 3D domain, with the same collocation points resolution as defined in the previous example. A fully connected neural network of four layers depth (three hidden layers) and layer width of 250 is defined. Instead of the gradient descent using all collocation points, a batch-gradient descent algorithm is used for RAM efficient optimization of the neural network weights with periodic random resampling of the training points after 100 iterations. A sinus activation function is used with Glorot uniform bias and weights initialization. The network is trained over 10000 iterations using the ADAM optimizer with a learning rate of 0.001, resulting in a test loss of 0.142.

\begin{figure}[ht]
  \centering
  \subfigure[Exact solution. \label{fig:subfig11}]{
    \begin{overpic}[width=0.28\linewidth,trim = 2.5cm 1.1cm 2.9cm 0,clip]{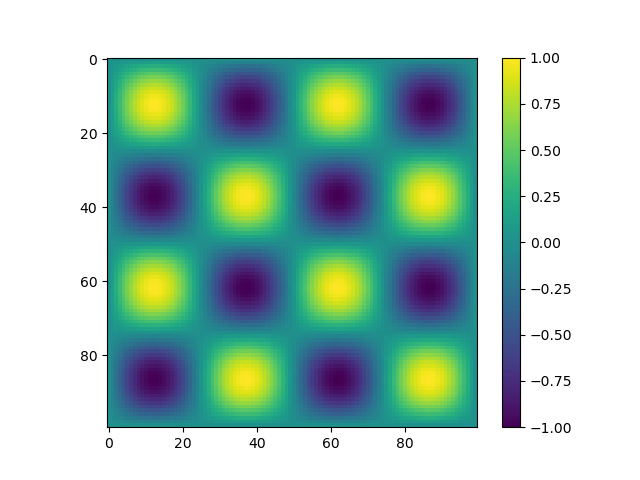}
 \put(-4,0){0}
\put(1,-6){0}
 \put(-4,84){1}
\put(85,-6){1}
\put(99,1){-1}
\put(99,42){~0}
\put(99,84){~1}
\end{overpic}
  }\hspace{1.0cm}
  \subfigure[PINN, constraint DBC.\label{fig:subfig21}]{
    \begin{overpic}[width=0.28\linewidth,trim = 2.5cm 1.1cm 2.9cm 0,clip]
{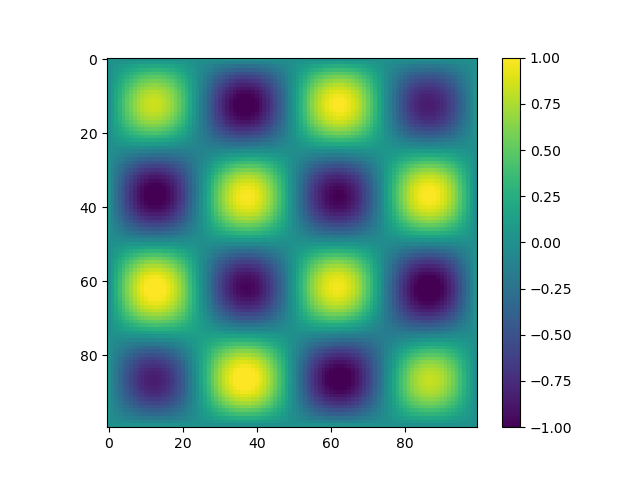}
\put(1,-6){0}
\put(85,-6){1}
\put(99,1){-1}
\put(99,42){~0}
\put(99,84){~1}
\end{overpic}
  }\hspace{1.0cm}
  \subfigure[PINN, DBC loss.\label{fig:subfig31}]{
    \begin{overpic}[width=0.28\linewidth,trim = 2.5cm 1.1cm 2.9cm 0,clip]
{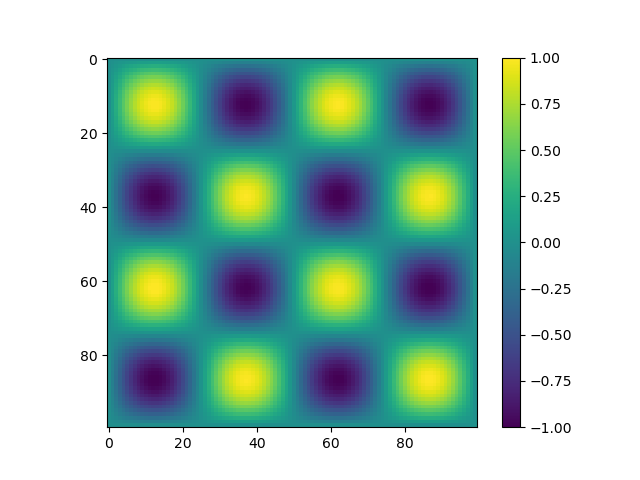}
\put(1,-6){0}
\put(85,-6){1}
\put(99,1){-1}
\put(99,42){~0}
\put(99,84){~1}
\end{overpic}
  }
  \caption{The real part of the acoustic pressure at $z=0.125$.}
  \label{fig:PINN_3D}
\end{figure}

In the case of enforcing the boundary conditions via the loss function using $\lambda_\mathrm{DBC}=100$, 20000 iterations (layer-width 180 have been made, resulting in a test loss of 0.0152. Figure \ref{fig:PINN_3D} shows the exact solution, the PINN solution with constraint DBC, and the PINN solution with DBC modeled as an additional loss term. The figure shows the acoustic pressure in a cut through the whole domain at $z=0.125$. It is observed that the number of iterations required rises with the number of loss terms and rises with the spatial dimensions. 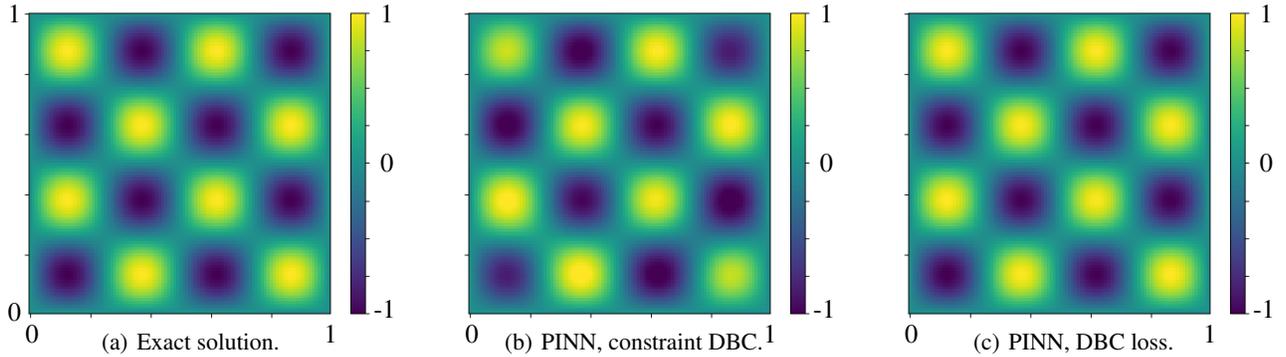

Until now, no feasible boundary conditions for room acoustic applications have been used. In room acoustic simulations, it is justified to use homogeneous Neumann boundary conditions as an approximation of the impedance jump between air and the room wall for low frequencies \cite{Kraxberger2023Validated}. As a next step, the loss-term modeled homogeneous Dirichlet boundary condition (sound soft) will be replaced by a homogeneous Neumann boundary condition (sound hard wall).

\subsection{Example 2a - Analytic solution in 3D, Room Acoustics} \label{sec:example-2a}
Following the previous example, the application of PINNs to solve the Helmholtz equation is extended to a 3D domain $\Omega = [0,1]^3$ with enforced Dirichlet boundary conditions. The governing equation is eq.~\eqref{eq:Helmholtz} with forcing of
\begin{equation}
    f(x,y,z) = 2 k^2 \sin(k x)\sin(k y)\sin(k z) \, ,
\end{equation}
with a wave number of $k = 2\pi/\lambda$ and wave length $\lambda=1/2$. This leads to an analytic solution of 
\begin{equation}
    p(x,y,z) = \sin(k x)\sin(k y)\sin(k z) \, .
\end{equation}
A homogeneous Neumann boundary condition $\nabla p(x,y,z) \cdot \bm n=0$ with $(x,y,z)\in \partial \Omega$ is considered by the loss function. 
The loss term is defined in eq.~\eqref{eq:loss-total}, with $\lambda_\mathrm{data}=\lambda_\mathrm{DBC}=0$, $\lambda_\mathrm{NBC}=5$, and $\lambda_\mathrm{PDE}=1$.
The PINN is set up using DeepXDE (PyTorch backend) with a layer-width of 180\footnote{The model coincides with the loss-term modeled Dirichlet boundary condition Helmholtz PINN from the previous example and can be seen as an evolution of this model.}. The network is trained over 30000 iterations by the ADAM optimizer with a learning rate of 0.001, resulting in a test loss of 0.125.

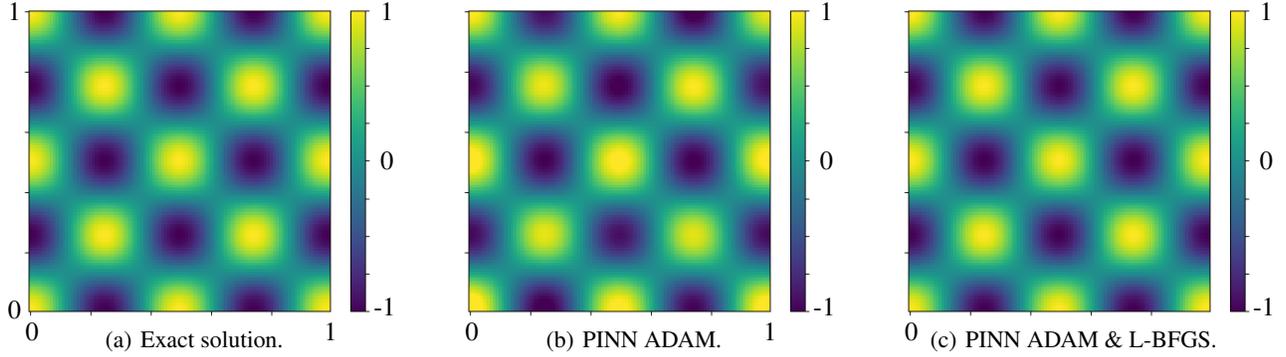
\begin{figure}[ht]
  \centering
  \subfigure[Exact solution.\label{fig:subfig1111}]{
  \begin{overpic}[width=0.28\linewidth,trim = 2.5cm 1.1cm 2.9cm 0,clip]{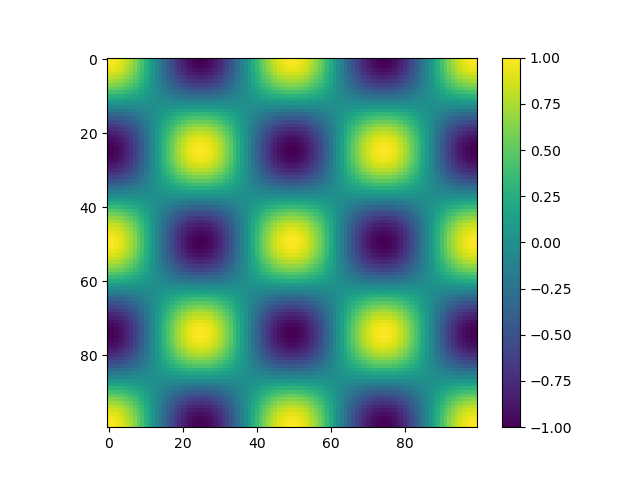}
 \put(-4,0){0}
\put(1,-6){0}
 \put(-4,84){1}
\put(85,-6){1}
\put(99,1){-1}
\put(99,42){~0}
\put(99,84){~1}
\end{overpic}
  }\hspace{1.0cm}
  \subfigure[PINN ADAM.\label{fig:subfig2111}]{
  \begin{overpic}[width=0.28\linewidth,trim = 2.5cm 1.1cm 2.9cm 0,clip]{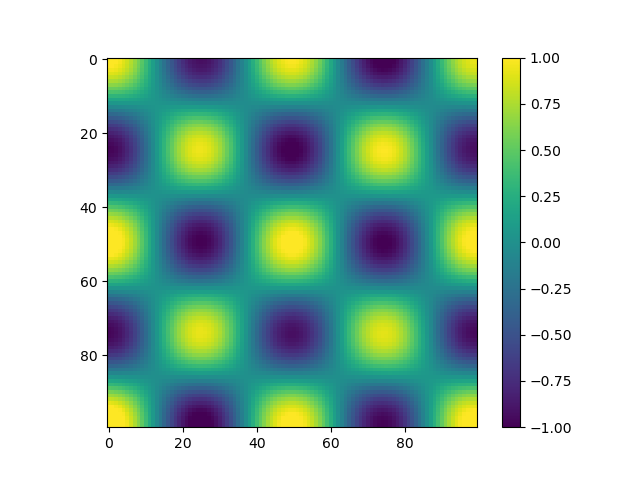}
\put(1,-6){0}
\put(85,-6){1}
\put(99,1){-1}
\put(99,42){~0}
\put(99,84){~1}
\end{overpic}
  }\hspace{1.0cm}
   \subfigure[PINN ADAM \& L-BFGS.\label{fig:subfig3111}]{
  \begin{overpic}[width=0.28\linewidth,trim = 2.5cm 1.1cm 2.9cm 0,clip]{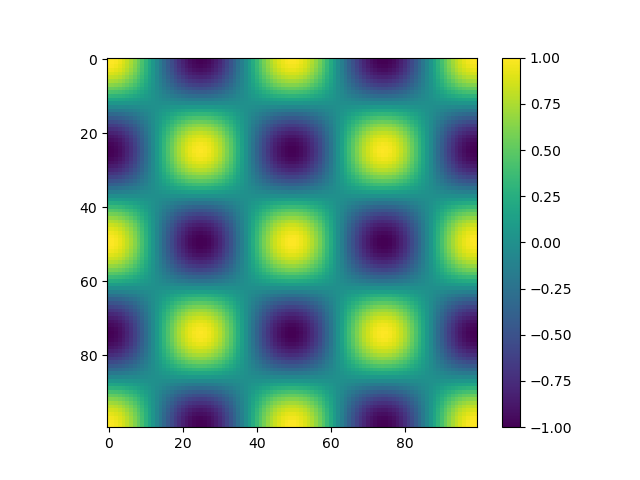}
\put(1,-6){0}
\put(99,1){-1}
\put(99,42){~0}
\put(99,84){~1}
\end{overpic}
  }
  \caption{The real part of the acoustic pressure at $z=0$ with NBC.}
  \label{fig:PINN_3D_Neumann}
\end{figure}

Figure \ref{fig:PINN_3D_Neumann} shows the exact solution, the PINN solution with constraint DBC, and the PINN solution with DBC modeled as an additional loss term. The figure shows the acoustic pressure in a cut through the whole domain at $z=0.125$. It is observed that the number of iterations required rises with the use of a Neumann boundary condition. Furthermore, iterating the final ADAM optimized network parameters with an L-BFGS optimizer for 15000 iterations brings down the test loss to 0.0142.

Another hyperparameter test was trying to predict both the real and the imaginary part of the acoustic pressure $p$, wherein the imaginary part is zero in the following example. As a consequence, when using the PINN for obtaining a complex acoustic pressure, it is suggested to train a network for the real and the imaginary part separately.

\subsection{Example 2b - Comparison to FEM in 3D, Room Acoustics} \label{sec:example-2b}
As a follow-up to the previous 3D room acoustic example, the source is modeled more realistically by a point source in the center of the room modeled by 
\begin{equation}
    f(x,y,z) =  2 k^2 \cos(k x)\cos(k y)\cos(k z) \mathrm{e}^{
    -\frac{(x-0.5)^2 + (y-0.5)^2 + (z-0.5)^2}{2\sigma^2}
    }    \, .
\end{equation}
The results for this excitation, are studied for several parameters $\sigma$, being a measure for the sharpness of the source distribution. For $\sigma\rightarrow0$, the source is very localized. In the current form, the source decreases it strength with $\sigma\rightarrow0$. Until a parameter $\sigma>0.1$, the original resolution of the field quantities was well enough to resolve the source distribution, leading to a condition $\lambda < 6\sigma$. As soon as this condition, does not hold, the resolution of the training data is too rough for the PINN to be accurate. For smaller $\sigma<0.1$, additional adaptive refinement was introduced, enriching the number of training points (recursively by five, for 10 iterations\footnote{\url{https://deepxde.readthedocs.io/en/latest/demos/pinn_forward/burgers.rar.html}}) in areas where the residual $r_\mathrm{PDE}$ were largest. After the enriching of training points, 3000 iterations were performed with the ADAM optimization following by 150000 iterations using the L-BFGS. With this strategy, $0.1>\sigma>0.02$ were predicted reasonable and a first step to resolve spatial and amplitude multi-scales of the problem. 


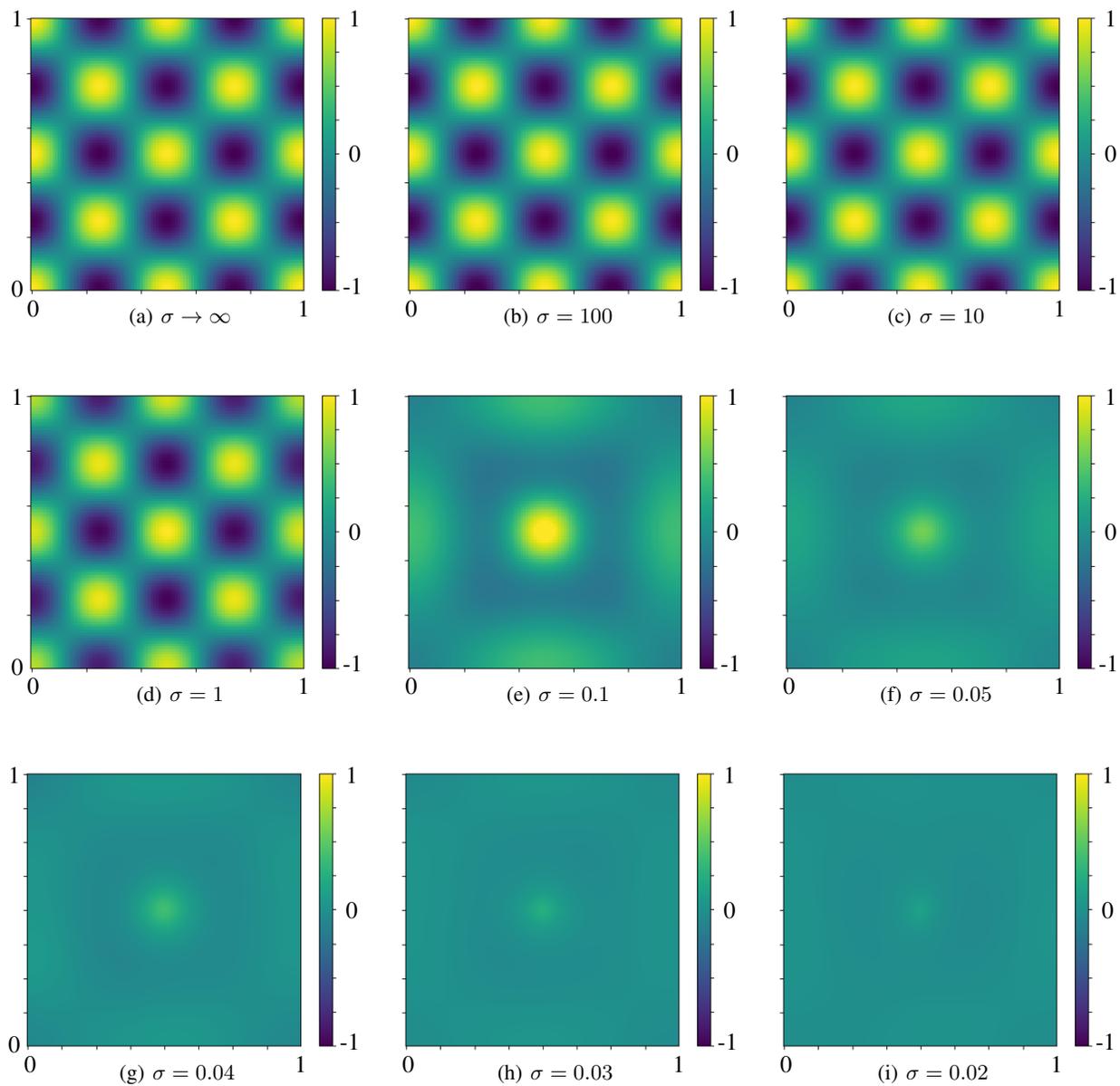
\begin{figure}[htp]
  \centering
  \subfigure[$\sigma\rightarrow\infty$\label{fig:subfig11111}]{
    \begin{overpic}[width=0.28\linewidth,trim = 2.5cm 1.1cm 2.9cm 0,clip]{media/Ex2b/plot_amplitude_noSigma.png}
 \put(-4,0){0}
\put(1,-6){0}
 \put(-4,84){1}
\put(85,-6){1}
\put(99,1){-1}
\put(99,42){~0}
\put(99,84){~1}
\end{overpic}
  }\hspace{0.7cm}
  \subfigure[$\sigma=100$\label{fig:subfig21111}]{
  \begin{overpic}[width=0.28\linewidth,trim = 2.5cm 1.1cm 2.9cm 0,clip]{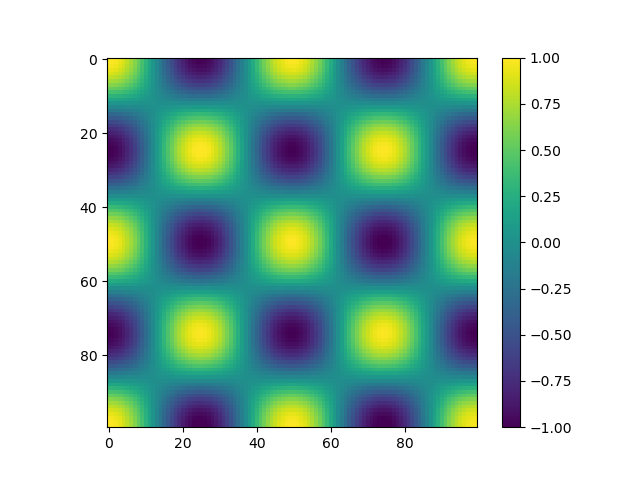}
\put(1,-6){0}
\put(85,-6){1}
\put(99,1){-1}
\put(99,42){~0}
\put(99,84){~1}
\end{overpic}
  }\hspace{0.7cm}
   \subfigure[$\sigma=10$\label{fig:subfig31111}]{
  \begin{overpic}[width=0.28\linewidth,trim = 2.5cm 1.1cm 2.9cm 0,clip]{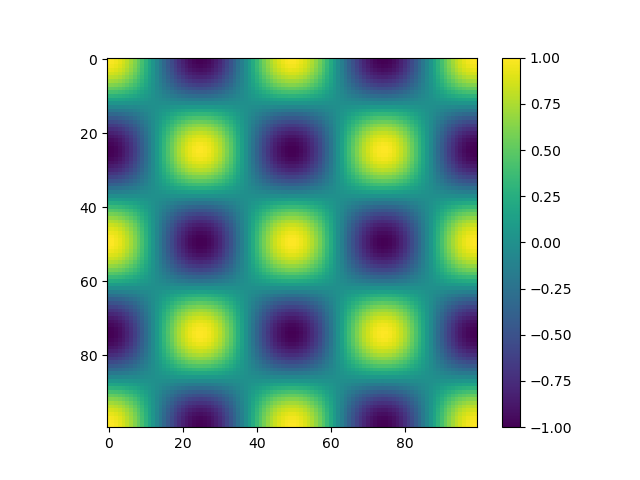}
\put(1,-6){0}
\put(85,-6){1}
\put(99,1){-1}
\put(99,42){~0}
\put(99,84){~1}
\end{overpic}
  }
    \subfigure[$\sigma=1$\label{fig:subfig11111b}]{
    \begin{overpic}[width=0.28\linewidth,trim = 2.5cm 1.1cm 2.9cm 0,clip]{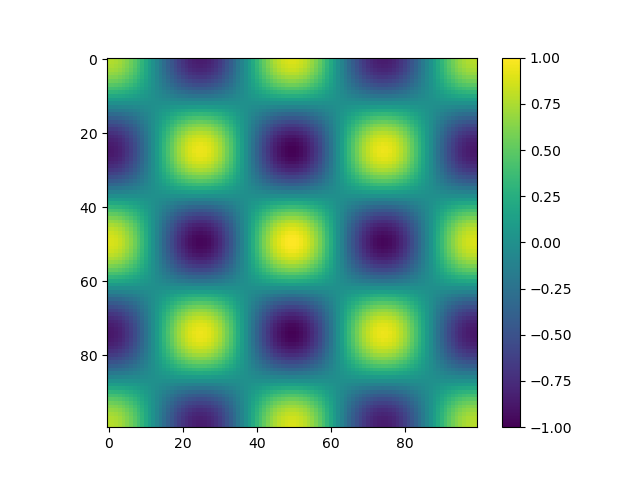}
 \put(-4,0){0}
\put(1,-6){0}
 \put(-4,84){1}
\put(85,-6){1}
\put(99,1){-1}
\put(99,42){~0}
\put(99,84){~1}
\end{overpic}
  }\hspace{0.7cm}
  \subfigure[$\sigma=0.1$\label{fig:subfig21111b}]{
  \begin{overpic}[width=0.28\linewidth,trim = 2.5cm 1.1cm 2.9cm 0,clip]{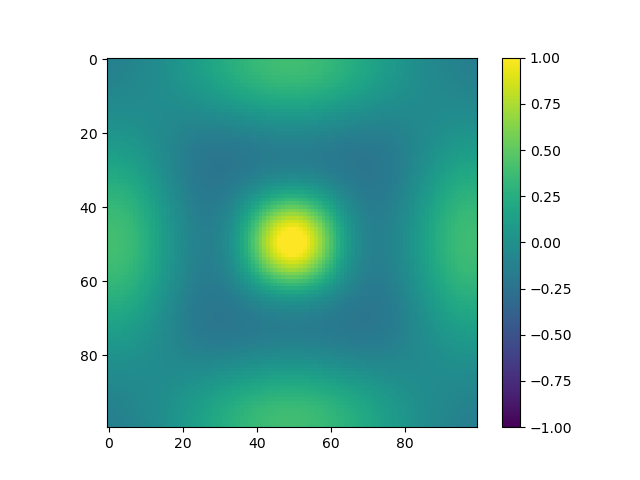}
\put(1,-6){0}
\put(85,-6){1}
\put(99,1){-1}
\put(99,42){~0}
\put(99,84){~1}
\end{overpic}
  }\hspace{0.7cm}
   \subfigure[$\sigma=0.05$\label{fig:subfig31111b}]{
  \begin{overpic}[width=0.28\linewidth,trim = 2.5cm 1.1cm 2.9cm 0,clip]{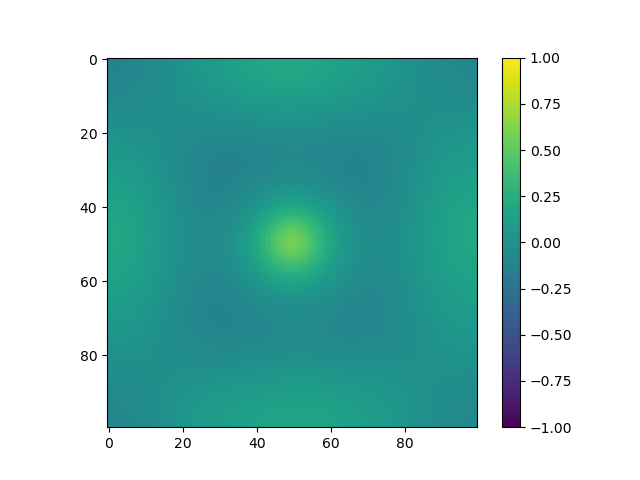}
\put(1,-6){0}
\put(85,-6){1}
\put(99,1){-1}
\put(99,42){~0}
\put(99,84){~1}
\end{overpic}
  }
      \subfigure[$\sigma=0.04$\label{fig:subfig11111c}]{
    \begin{overpic}[width=0.28\linewidth,trim = 2.5cm 1.1cm 2.9cm 0,clip]{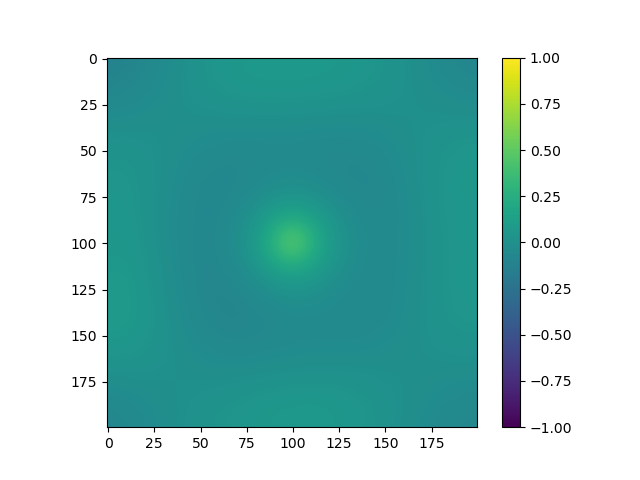}
 \put(-4,0){0}
\put(1,-6){0}
 \put(-4,84){1}
\put(85,-6){1}
\put(99,1){-1}
\put(99,42){~0}
\put(99,84){~1}
\end{overpic}
  }\hspace{0.7cm}
  \subfigure[$\sigma=0.03$\label{fig:subfig21111c}]{
  \begin{overpic}[width=0.28\linewidth,trim = 2.5cm 1.1cm 2.9cm 0,clip]{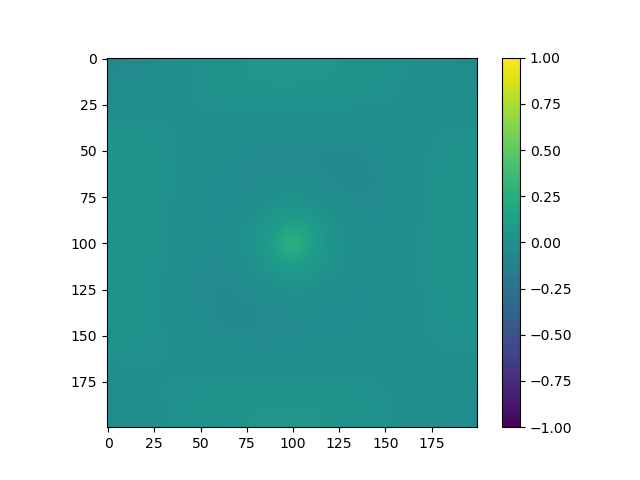}
\put(1,-6){0}
\put(85,-6){1}
\put(99,1){-1}
\put(99,42){~0}
\put(99,84){~1}
\end{overpic}
  }\hspace{0.7cm}
   \subfigure[$\sigma=0.02$\label{fig:subfig31111c}]{
  \begin{overpic}[width=0.28\linewidth,trim = 2.5cm 1.1cm 2.9cm 0,clip]{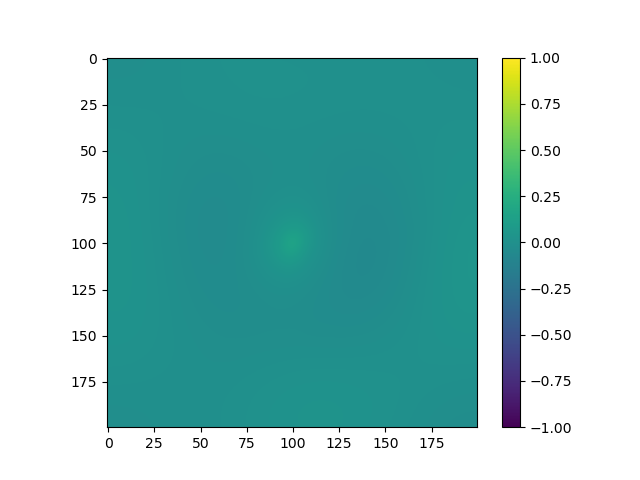}
\put(1,-6){0}
\put(85,-6){1}
\put(99,1){-1}
\put(99,42){~0}
\put(99,84){~1}
\end{overpic}
  }
  \caption{The real part of the acoustic pressure at $z=0.5$.}
  \label{fig:Sigma}
\end{figure}

\begin{figure}[htp]
  \centering
  \subfigure[Sparse FEM mesh (FEM-sparse).]{
    \includegraphics[width=0.3\linewidth]{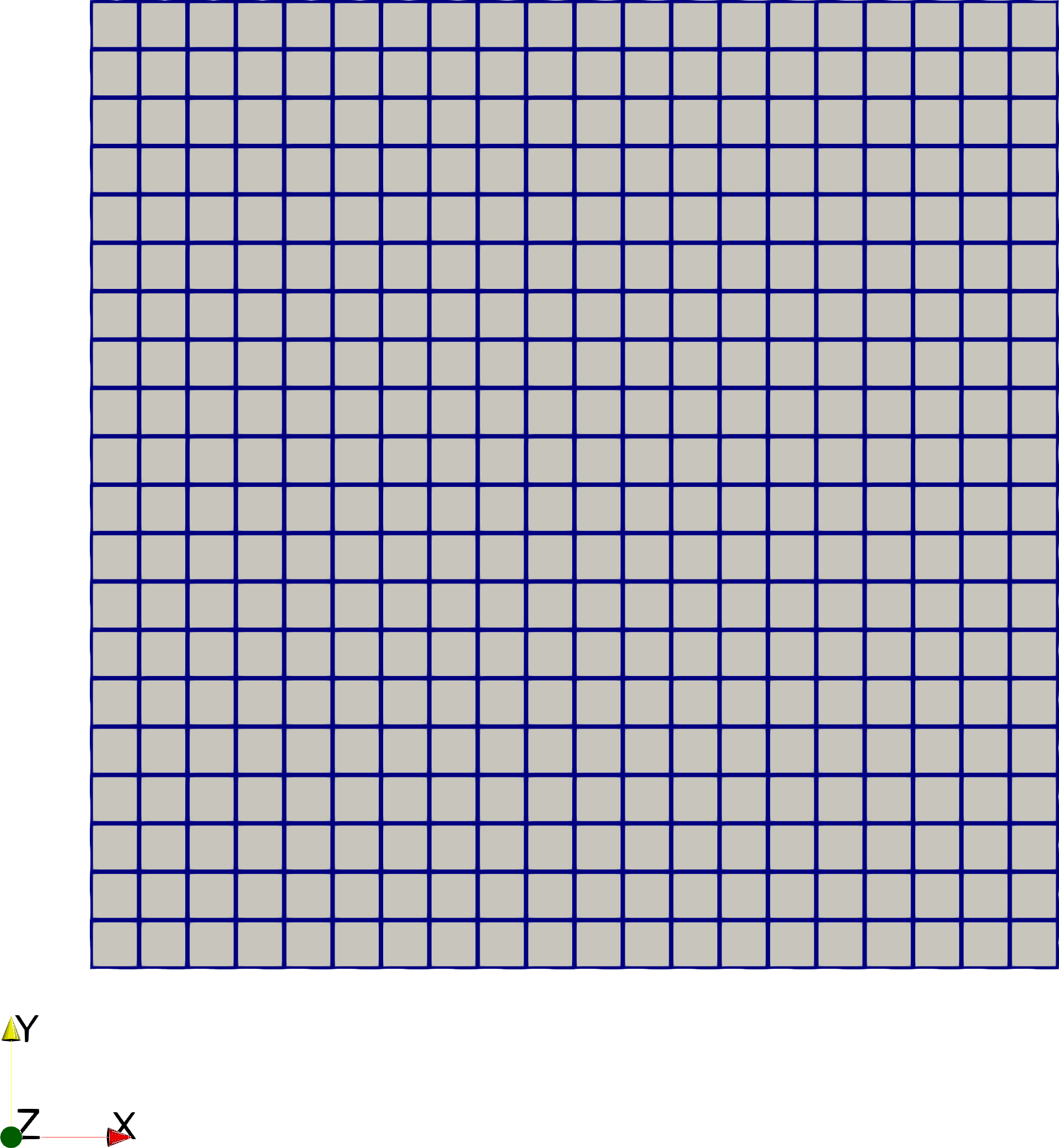}
  }
  \hspace{3cm}
  \subfigure[Fine FEM mesh (FEM-fine).]{
    \includegraphics[width=0.3\linewidth]{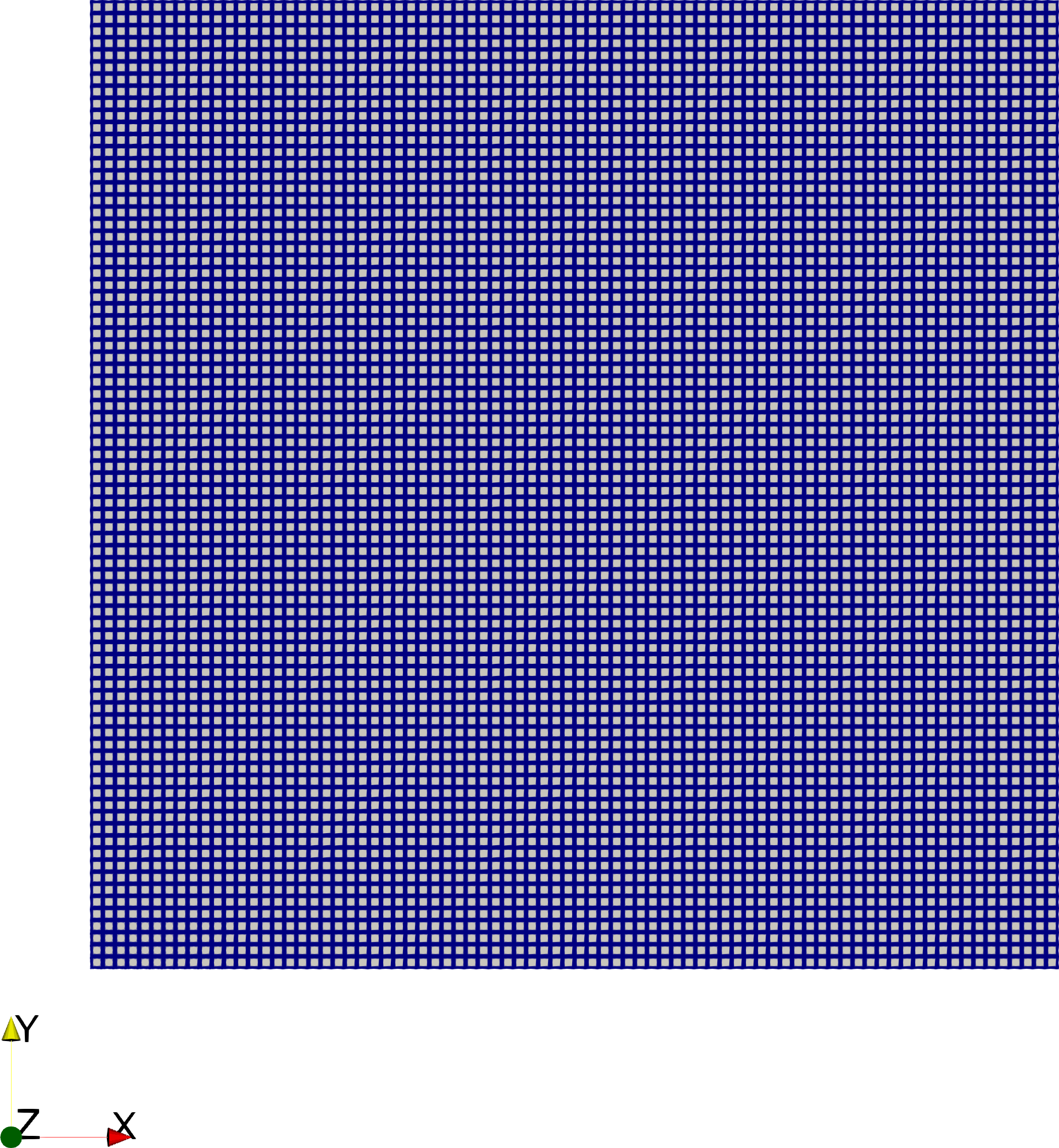}
  }
  \caption{Cross-section through the sparse and fine meshes used for the FEM computations at $z=0.5$.}
  \label{fig:FEM-meshes}
\end{figure}

The FEM reference solution based on the same setup was carried out in openCFS \cite{Schoder2022openCFS}. Thereby, a sparse and a fine mesh are used as depicted in fig.~\ref{fig:FEM-meshes}. The sparse mesh discretizes the cube's edge by 20 linear elements (approximately 10 degrees of freedom per wave length) , resulting in a total number of 8000 elements, and the fine mesh discretizes the cube's edge by 80 linear elements (approximately 40 degrees of freedom per wave length), resulting in 512000 elements of the computational domain.

A relative error measure $Err_\mathrm{rel}$ is evaluated, such that
\begin{equation}
    Err_\mathrm{rel} = \sqrt{\frac{
                         \sum_{i=1}^{N_\mathrm{pts}} \left( p_{i,\mathrm{ref}} - p_{i} \right)^2}
                        {\sum_{i=1}^{N_\mathrm{pts}} \left( p_{i,\mathrm{ref}} \right)^2} }  \, ,
\end{equation}
where $N_\mathrm{pts}$ is the number of evaluation points, $p_{i,\mathrm{ref}}$ is the reference pressure and $p_{i}$ is the pressure for which the agreement against the reference pressure should be quantified. The $N_\mathrm{pts}=10000$ evaluation points have been placed to the locations where the PINN prediction is evaluated, i.e., a grid of $\SI{10}{\centi\meter} \times \SI{10}{\centi\meter}$ across the computational domain at $z=\SI{0.5}{\meter}$. To interpolate the FEM solution to this grid, the FE basis functions are evaluated at the according location in the respective element, as implemented in openCFS \cite{Schoder2022openCFS}. The errors are denoted as follows:
$Err_\mathrm{rel}^{\text{PINN,FEMsparse}}$ is the error between the PINN solution and the FEM solution obtained with the sparse mesh, 
$Err_\mathrm{rel}^{\text{PINN,FEMfine}}$ is the error between the PINN solution and the FEM solution obtained with the fine mesh, and
$Err_\mathrm{rel}^{\text{FEM}}$ is the error between the FEM solutions obtained with fine and sparse meshes.

\begin{table}[H]
\centering
\caption{Performance comparison between PINN and FEM. For FEM-sparse and FEM-fine the given durations are the CPU-hours, and for PINN the duration denotes the time on one GPU.}
\begin{tabularx}{\textwidth}{@{}rccccccc@{}}
\toprule
\multirow{2}{*}[-\aboverulesep]{$\sigma$} & \multicolumn{4}{c}{Duration (mm:ss)} & \multicolumn{3}{c}{Error}   \\
\cmidrule(lr){2-5} \cmidrule(lr){6-8}
 & PINN-training & PINN-prediction & FEM-sparse & FEM-fine & $Err_\mathrm{rel}^{\text{PINN,FEMsparse}}$ & $Err_\mathrm{rel}^{\text{PINN,FEMfine}}$ & $Err_\mathrm{rel}^{\text{FEM}}$ \\
\midrule        
$0.1$   & 2374:40.4 & 00:00.05 & 00:16.8 & 17:44.8 & $0.5754$ & $0.9718$ & $0.9275$ \\
$1.0$   & 2512:01.8 & 00:00.05 & 00:17.6 & 17:18.8 & $0.0454$ & $0.0997$ & $0.0924$ \\
$10.0$  & 2280:20.0 & 00:00.05 & 00:20.9 & 19:00.0 & $0.0300$ & $0.0243$ & $0.0352$ \\
$100.0$ & 2565:38.8 & 00:00.05 & 00:20.7 & 19:19.5 & $0.0303$ & $0.0249$ & $0.0352$ \\
\bottomrule
\end{tabularx}
\label{tab:performance}
\end{table}



\section{Conclusion} \label{sec:conclusion}

In this working paper, the potential of using PINNs to approximate forward solutions of the Helmholtz equation is demonstrated for geometrically simple 2D and 3D problems (square or cubic computational domains) with homogeneous Neumann and Dirichlet boundary conditions. Thereby, the readily available PINN framework deepXDE is used to implement the neural networks and their respective cost functions. Having very low training losses and excellent agreement with the analytical solutions, the results exhibit a promising ability of PINNs as forward solvers.

Regarding computational cost, Tab.~\ref{tab:performance} exhibits an interesting behavior: Using the trained PINN as a forward-pass solver for the PDE, it is orders of magnitude faster than using a FEM solver. This highlights the potential of trained PINNs as surrogate models within optimization frameworks. However, it has to be noted that a study regarding the generalization abilities of the PINN is required prior to using the PINN in an optimization framework.

Future work may include a hyperparameter optimization of the PINNs in order to achieve a smaller test loss. Additionally, it is subject to further research, how well the PINN approach generalizes to more complex geometries to be applicable in real-world problems, i.e. large complex non-cuboid shapes. Nevertheless, the results presented in this working paper encourage future work.

\section*{Data and code availability}
 The data and code is available on reasonable request from the authors.


\bibliographystyle{unsrtDOI.bst}  
\bibliography{references}

\end{document}